# Anomalous Raman Response in 2D Magnetic FeTe under Uniaxial Strain: Tetragonal and Hexagonal Polymorphs

Wuxiao Han,[1,2] Tiansong Zhang,[1,2] Pengcheng Zhao,[1,2] Longfei Yang,[2] Mo Cheng,[3] Jianping Shi,[3] and Yabin Chen[1,2,4,*]

[1]*Advanced Research Institute of Multidisciplinary Sciences, Beijing Institute of Technology (ARIMS), Beijing 100081, China.*

[2]*School of Aerospace Engineering, Beijing Institute of Technology, Beijing 100081, China.*

[3]*The Institute for Advanced Studies, Wuhan University, Wuhan 430072, China.*

[4]*BIT Chongqing Institute of Microelectronics and Microsystems, Chongqing, 400030, China.*

[*]Correspondence and requests for materials should be addressed to Prof. Yabin Chen: chyb0422@bit.edu.cn

## ABSTRACT

Two-dimensional (2D) Fe-chalcogenides have emerged with rich structures, magnetisms and superconductivities, which sparked the growing research interests in the torturous transition mechanism and tunable properties for their potential applications in nanoelectronics. Uniaxial strain can produce a lattice distortion to study symmetry breaking induced exotic properties in 2D magnets. Herein, the anomalous Raman spectrum of 2D tetragonal (*t*-) and hexagonal (*h*-) FeTe were systematically investigated via uniaxial strain engineering strategy. We found that both *t*- and *h*-FeTe keep the structural stability under different uniaxial tensile or compressive strain up to ± 0.4%. Intriguingly, the lattice vibrations along both in-plane and out-of-plane directions exceptionally hardened (softened) under tensile (compressive) strain, distinguished from the behaviors of many conventional 2D systems. Furthermore, the difference in thickness-dependent strain effect can be well explained by their structural discrepancy between two polymorphs of FeTe. Our results could provide a unique platform to elaborate the vibrational properties of many novel 2D materials.

# INTRODUCTION

Two-dimensional (2D) transition metal dichalcogenides (TMDs) exhibit exotic physical properties, including their rich magnetism,[1,2] structural phase transition,[3,4] unconventional superconductivity and charge density waves,[5-8] making them an extensive research platform in the fields of spin electronics and memory nanodevices.[9,10] The interplay among lattice, magnetism and superconductivity is a common theme across iron-based superconductors (*e.g.*, FeS, FeSe, and FeTe).[11-13] The geometric structures of 2D magnetic materials enable their lattices and spin states to be effectively controlled by external stimuli, such as hydrostatic pressure[14] and strain.[15] Indeed, pressure-induced tetragonal (*t*-) to hexagonal (*h*-) phase transitions in FeS caused the decrement and eventual vanishing of superconducting transition temperature, as pressure increased up to ~4 GPa.[16,17] More interestingly, the superconductivity in FeSe nanofilms can be suppressed by the tensile strain, whereas the superconductivity reappeared in monolayer FeSe when its pronounced spin density waves were suppressed under external strain.[18,19] With the applied strain, it was found that the magnetic order and superconductivity of FeSe were significantly sensitive to its lattice symmetry, indicating their complicated and ambiguous relationships.[20] Notably, it was predicted that the intrinsic FeTe may exhibit the highest superconducting temperature among iron chalcogenides, owing to its unique structure and spin state,[21-23] which needs to be proved in experiments, despite several studies tentatively performed on the doped FeTe.[24-27] Very recently, both 2D layered antiferromagnetic *t*-FeTe and non-layered ferromagnetic *h*-FeTe nanosheets were controllably synthesized by phase-tunable chemical vapor deposition (CVD) approach,[28-30] while the investigations on their strain-modulated lattice structures and phonon properties are still lacking.

Strain engineering, as an efficient approach, has been widely utilized to tune the lattice structure and thus regulate the phonon-electron coupling in strongly correlated systems.[31,32] Uniaxial strain can controllably produce a continuous lattice distortion to break its innate symmetry,[33] and theoretical calculations have predicted a strong influence of elastic strain on the ground-state electronic properties as well as magnetic orders in the 2D magnets.[34-36] For instance, the unique ($\pi$, $\pi$)-charge-ordered state of $Fe_{1.1}Te$ can be stabilized through the applied strain, due to a significant monoclinic distortion, which evidently demonstrated that magnetic order was strongly correlated with the superconducting state.[37] Meanwhile, the previous studies on *t*-FeTe immaturely suggested the magnetic transition from low-pressure antiferromagnetic to high-pressure ferromagnetic order under hydrostatic pressure.[38-40] Therefore, the experimental explorations on lattice structure and vibration of *t*-FeTe and *h*-FeTe under uniaxial strain may provide very meaningful insights into its effect on their exotic properties.

2D FeTe nanosheets with the tunable phases, thickness and magnetic orders are extremely rare in TMDs family,[28-30] offering a desirable platform to investigate their lattice dynamics as well as the attractive mechanism in iron-based superconductors. In this work, we investigated the structural and vibrational properties of 2D *t*-FeTe and *h*-FeTe nanosheets under uniaxial strain. It is found that both *t*- and *h*-FeTe keep structural stability under tensile (compressive) strain up to +(-)0.4%, confirmed by *in situ* Raman spectroscopy. In-plane and out-of-plane vibrations were highly sensitive to the uniaxial strain along in-plane direction. Interestingly, both $E_g$ and $A_{1g}$ modes linearly hardened and softened under tensile and compressive strains, respectively. This anomalous phenomenon was attributed to the strong spin-phonon coupling, well consistent with other 2D magnetic materials.[33,41] Thickness-dependent Raman shifting rate was individually related to the lattice symmetry of FeTe. Our results could shed light to explore more novel properties and further comprehend physical mechanism of numerous 2D materials.

**RESULTS AND DISCUSSION**

The schematics of straining device and lattice structures of 2D *t*-FeTe and *h*-FeTe crystals are shown in Figure 1a. The 2D FeTe nanoflakes with two distinct phases were grown by using temperature-mediated CVD approach as reported in our previous study.[29] The homemade four-point bending setup[42,43] was used to apply the tunable and uniaxial strain along the in-plane direction of FeTe, where the nanoflake specimens were located at the center of the flexible polyethylene terephthalate (PET) substrate (Figure 1a). A thin polymethyl methacrylate (PMMA) film was coated on the surface of FeTe and PET, in order to protect FeTe samples from any potential environmental contaminations, degradation, and slippage against the PET substrate during loading tests. Meanwhile, this versatile setup can effectively realize the tensile and compressive states by simply flipping the entire PET substrate over. It was confirmed that the as-fabricated strain device together with FeTe were virtually strain-free before bending. Therefore, the applied strain $\varepsilon$ can be derived as $\varepsilon = \tau/2R$ upon bending, where $R$ and $\tau$ denotes the curvature radius and thickness of PET substrate, respectively. In our case, $\tau$ was typically around 200 μm. Importantly, our homemade straining device is compatible with the external optical spectroscopy, allowing us to carry out strain-dependent Raman characterizations under various bending conditions, as illustrated in Figure 1a and S1.

To facilitate understanding lattice structures of *t*- and *h*-FeTe nanoflakes, the systematic structural characterizations were performed by using atomic force microscopy (AFM) and high-resolution transmission electron microscopy (HRTEM), as shown in Figure 1b-g. Obviously, *t*-FeTe possesses a layered structure with P4/nmm space group, in which Fe atoms are distributed between the double slabs of Te atomic layers along the interlayer direction. In comparison, *h*-FeTe belongs to $P6_3$/mmc space group with a non-layered structure. The AFM results proved that the extracted thickness was around 10.6 nm for *t*-FeTe (Figure 1b) and 11.8 nm for *h*-FeTe (Figure 1e). The clean surfaces of 2D FeTe flakes on the flexible PET substrates were free of any cracks or pre-strain induced wrinkles. More AFM characterizations of 2D FeTe nanoflakes with different thicknesses were displayed in Figure S2. Intriguingly, 2D FeTe nanoflakes obviously exhibited two distinct morphologies, 90º in *t*-FeTe and 120º (60º) in *h*-FeTe due to their own lattice symmetry, enabling one to intuitively distinguish the phase structure of any given FeTe nanoflakes after CVD growth. Figure 1c and 1f exhibit the obtained HRTEM results with atomic resolution, carried out along [001] zone axis. The extracted lattice constant 3.8 Å of *h*-FeTe was relatively smaller than that (3.9 Å) of *t*-FeTe as indicated by the yellow dashed lines, well consistent with our previous results.[14,29] The atomic arrangements of 2D FeTe perfectly comply with their specific lattice symmetries, which were further verified by selected area electron diffraction (SAED) measurements. The SAED patterns in Figure 1d and 1g present the tetragonal and hexagonal symmetry, respectively, and the corresponding lattice constants were in good agreement on HRTEM results. Meanwhile, the bright and sharp diffraction spots indicate the uniform and crystalline quality of our 2D FeTe nanosheets.

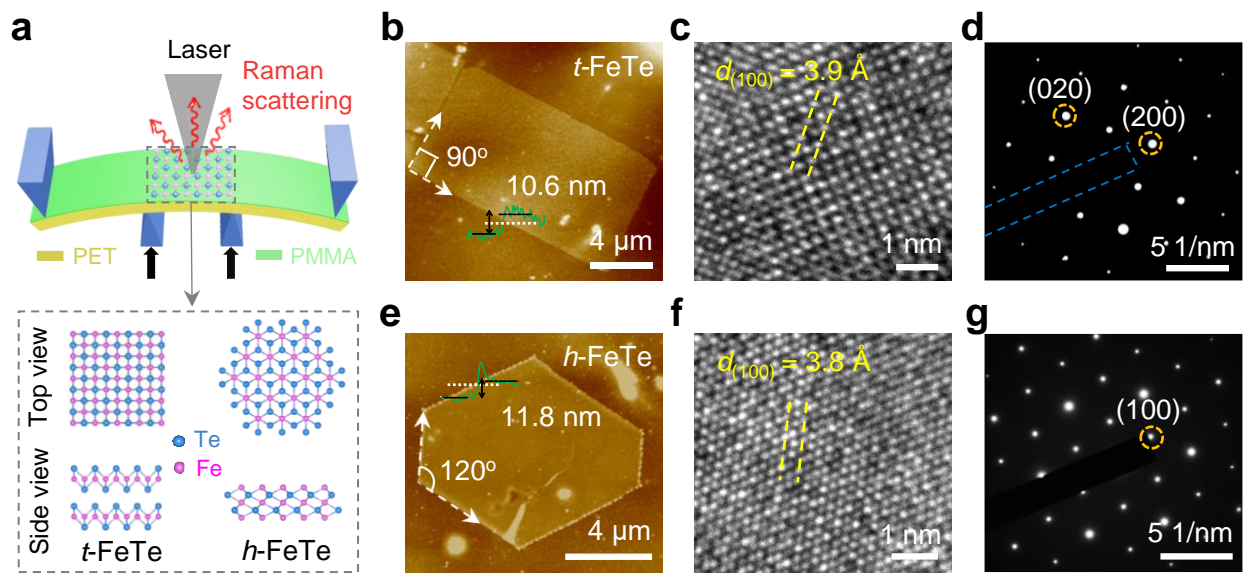

**Figure 1. Schematic of the straining system and structural characterizations of 2D crystalline *t*-FeTe and *h*-FeTe nanosheets. a)** Schematic of four-point bending setup used to strain FeTe on a flexible PET substrate. The *t*-FeTe with a layered structure belongs to P4/nmm space group, in which Fe atomic layer is distributed between the double slabs of Te atomic layers in the interlayer direction. The *h*-FeTe with a non-layered structure is with $P6_3/mmc$ space group. **b, e)** AFM images of *t*-FeTe nanosheet with a thickness of 10.6 nm (b) and *h*-FeTe nanosheet with a thickness of 11.8 nm (e). **c, f)** HRTEM images of *t*-FeTe (c) and *h*-FeTe (f) nanosheets, and their distances between (100) planes are determined as 3.9 and 3.8 Å, respectively. **d, g)** The corresponding SAED patterns of *t*-FeTe (d) and *h*-FeTe (g) nanoflakes. The sharp diffraction spots indicate high quality of the crystalline FeTe samples. The blue dash box depicts the position of TEM baffle which obscures partial diffraction points.

To explore the physical insights into the strain effect on lattice vibrations of 2D FeTe, we performed *in situ* Raman measurements on both *t*-FeTe and *h*-FeTe under tunable uniaxial strain. The maximum tensile and compressive strain was up to ±0.4%. 2D FeTe nanosheets were transferred onto the surface of PET substrate by a polystyrene-assisted method, as developed in our previous research.[29] AFM characterizations of the pre-transferred 2D FeTe nanoflakes were shown in Figure S2. The representative Raman spectrum of a *t*-FeTe nanosheet (10.6 nm thick) under uniaxial strain is presented in Figure 2a. The Raman peaks at ~122 and 140 cm$^{-1}$ of *t*-FeTe can be assigned to $E_g$ and $A_{1g}$ modes, which vibrates along in-plane and out-of-plane directions according to group theory, respectively. The Raman shifts of $E_g$ mode displayed the apparent blueshifts (redshifts) under tensile (compressive) strains, owing to the softened phonon vibrations. As shown in Figure 2b, the extracted $E_g$ frequencies by Lorentz fittings increased monotonically from negative (compression) to positive (tension) strains. Moreover, the strain-induced Raman response of 2D *t*-FeTe nanoflakes with different thicknesses (from ~9.3 to 41.8 nm) exhibited the reproducible tendency (more details in Supplementary Figures S3). Similarly, $A_{1g}$ mode of *t*-FeTe nanoflakes showed the high sensitivity to the external strain along in-plane direction, accompanied with the dramatically enhanced frequencies from compression to tension states, as shown in Figure 2c. This phenomenon can be well explained by the reduced interlayer spacing, resulted from the intralayer stretching due to its positive Poisson's ratio of *t*-FeTe.[44]

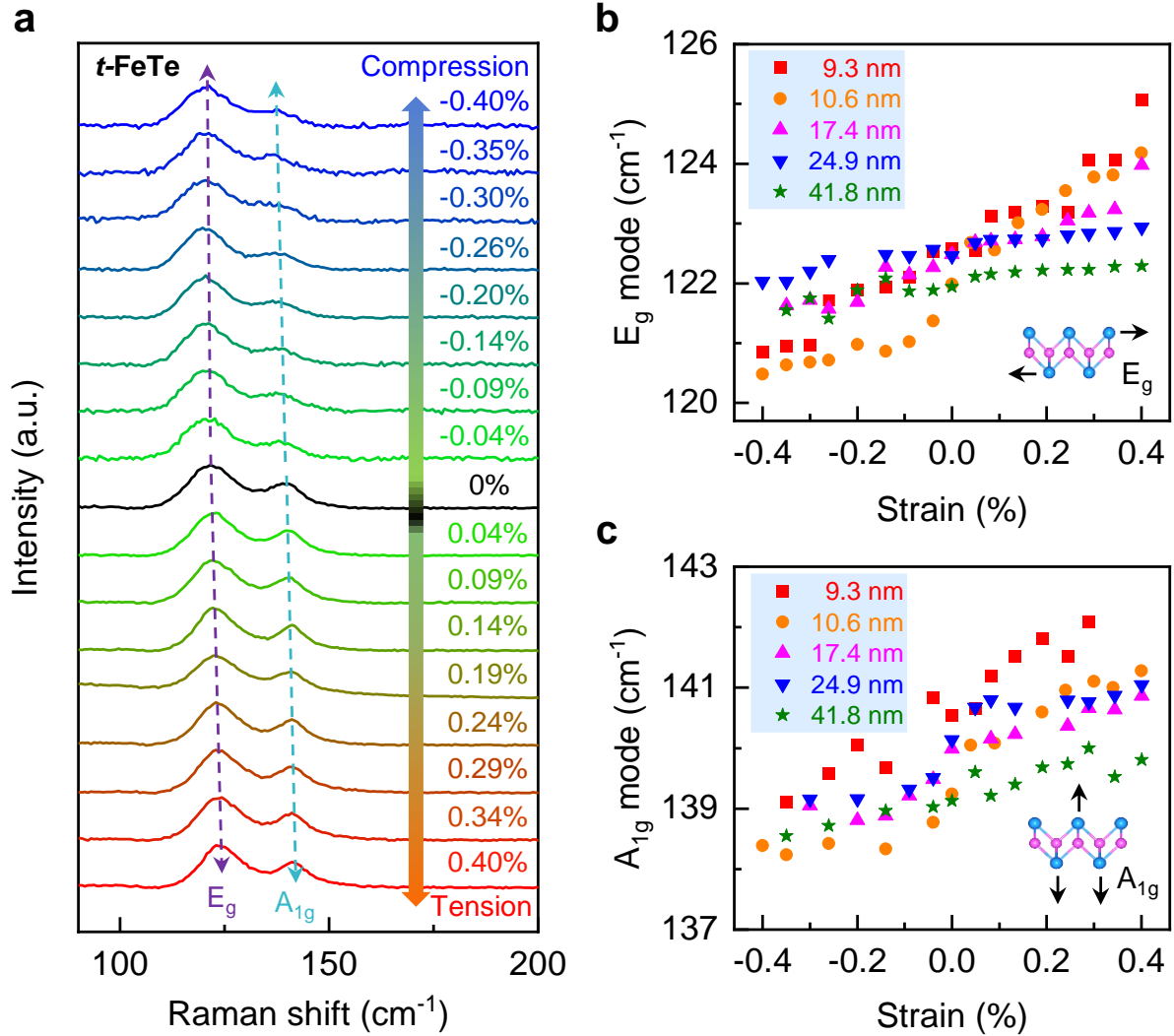

**Figure 2. Raman spectrum of 2D *t*-FeTe with different thicknesses under uniaxial strain. a)** Raman spectrum evolution of a 10.6 nm-thick *t*-FeTe when the tensile and compressive strains were applied along in-plane direction. The dash lines with arrows are guides for the eyes. **b)** Raman shift of $E_g$ mode of *t*-FeTe with different thicknesses from 9.3 to 41.8 nm. **c)** Raman shift of $A_{1g}$ mode of *t*-FeTe with different thicknesses. The insert shows the schematic diagram of crystalline structure and two typical Raman modes of 2D *t*-FeTe.

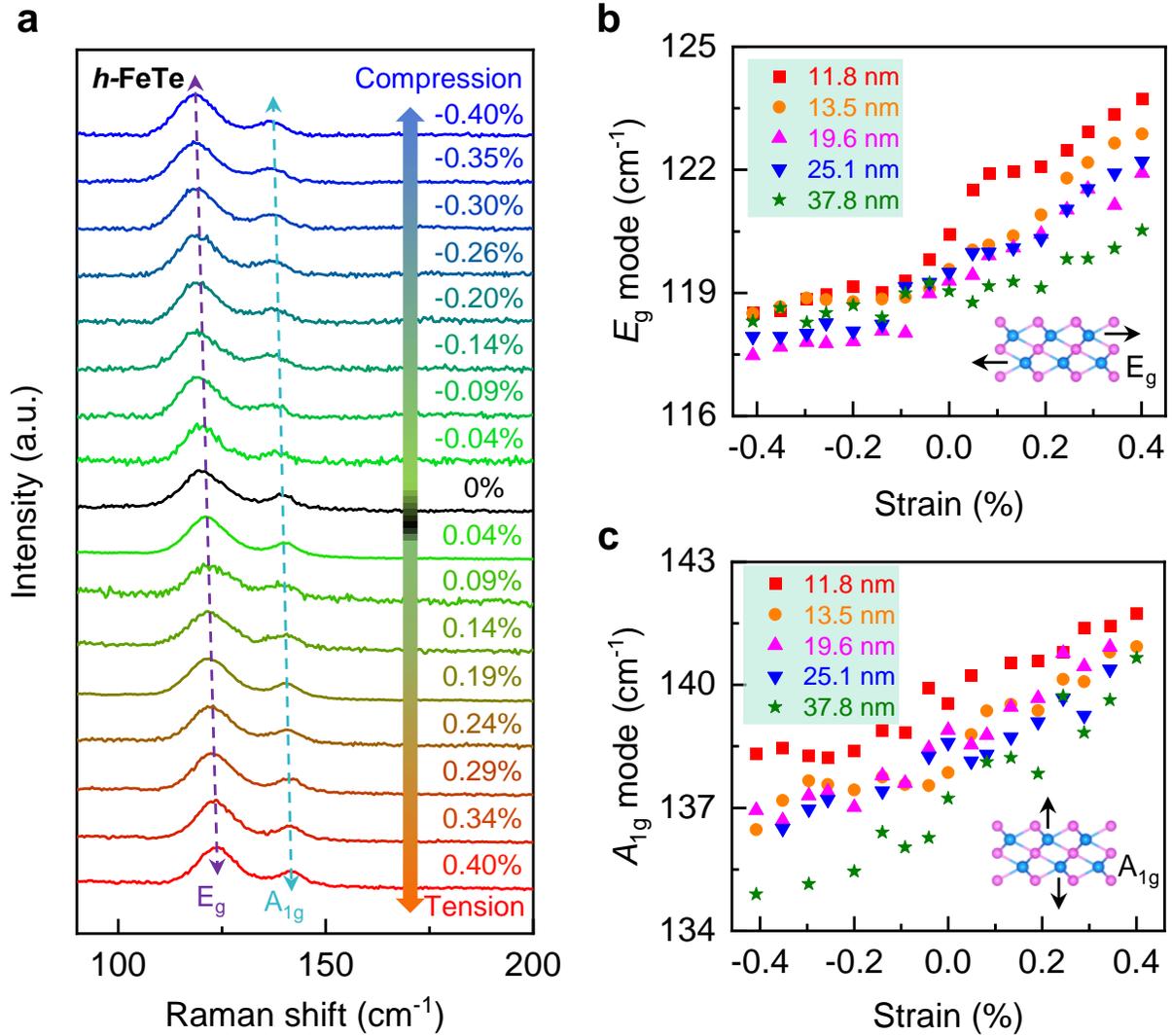

**Figure 3. Raman spectroscopy characterizations of 2D *h*-FeTe with different thicknesses under uniaxial strain. a)** Raman spectrum evolution of a 11.8 nm-thick *h*-FeTe under the tensile and compressive strains applied along its in-plane direction. The dash lines with arrows are used as guide for the eyes. **b)** Raman shift of $E_g$ mode of *h*-FeTe with various thicknesses, ranging from 11.8 to 37.8 nm. **c)** Raman shift of $A_{1g}$ mode of *h*-FeTe with different thicknesses. The insert shows the schematic diagram of atomic structures of 2D *h*-FeTe together with its typical Raman modes.

In comparison, we further carried out the strain-modulated Raman measurements of 2D *h*-FeTe. The representative results based on a 11.8 nm thick nanoflake are presented in Figure 3a.

It is obvious that the Raman response of *h*-FeTe under strain was comparable with that of *t*-FeTe, indicating that their individual lattice symmetry was retained under strain even up to ~4.0%. Furthermore, we acquired strain-dependent Raman spectra of many *h*-FeTe nanoflakes with the variable thicknesses from ~11.8 to 37.8 nm, and their Raman results showed the consistent results as well (more details in Supplementary Figures S4). In Figure 3b, by Lorentz fitting of each Raman curve, the obtained in-plane $E_g$ mode became remarkably hardened (softened) under tensile (compressive) strains. Both $E_g$ and $A_{1g}$ frequencies exhibited the significant sensitivities to the in-plane strain with a quasi-linear relationship with the variable strains. Notably, the positive slopes d$\omega$/d$\varepsilon$ of $E_g$ and $A_{1g}$ modes in *t*-FeTe and *h*-FeTe are dramatically different from many conventional 2D layered materials,[45] and this anomalous Raman response will be discussed later (Figure 4).

To interpret the unexpected Raman behaviors of 2D FeTe nanosheets induced by strain, we have first summarized the Raman response of many different 2D materials under strain. As shown in Figure 4, the Raman modes of 2D materials such as $MoS_2$, $WS_2$, $WSe_2$ and $VS_2$ exhibit a clear negative slope in both in-plane and out-of-plane vibrations under uniaxial tensile strain.[46-49] Due to the weakening of restoring force, their Raman modes gradually soften and the energy of the corresponding phonon modes reasonably decreased with the applied strain. In comparison, the in-plane and out-of-plane Raman shifts of b-As presented different slopes under an uniaxial tensile strain along armchair direction, in which negative Poisson's ratio played the critical role.[50] However, the remained materials with stronger magnetic orders exhibited remarkably different Raman responses under strain, including $Cr_2S_3$, GeSe and $TaS_2$.[33,51,52] For $Cr_2S_3$, both in-plane and out-of-plane Raman vibrations displayed a redshift under compressive strain, which was closely related to its strong electron-phonon coupling.[33] Similarly, this phenomenon also occurred in $TaS_2$ with the apparently blue-shifted Raman mode under tensile strain.[52] The $B_{3g}$ and $A_g^3$ modes of GeSe showed blueshift when the uniaxial tensile strain applied along its armchair direction, due to the pressure-induced variations of bond angles.[51] For 2D FeTe, it is noted that both $E_g$ and $A_{1g}$ modes primarily originate from the vibration of Te atoms.[53,54] In our case, the distinct responses of FeTe under uniaxial strain can be tentatively attributed to two aspects: (1) FeTe exhibits complex magnetic orders under strain, resulting in the remarkably stronger electron-phonon coupling which was confirmed in other typical telluride materials previously;[55-57] (2) FeTe can easily shrink its out-of-plane lattice constant once in-plane strain was applied. The recent experimental studies on FeTe and FeSe have evidently proved their weak elastic modulus, leading to more easily compression along out of plane direction due to its van der Waals characteristic.[58-60] This scenario is well consistent with literature,[27] in which Fe-Te-Fe bond angles increased dramatically due to the out-of-plane lattice contraction.

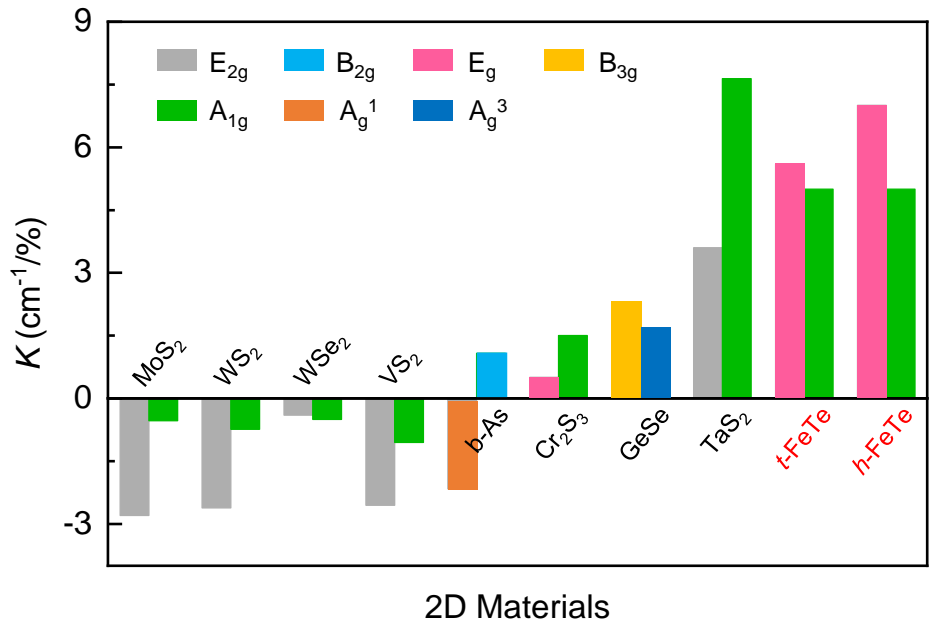

**Figure 4. The Raman shifting rates $K$ of 2D FeTe, compared with numerous conventional 2D materials under uniaxial strain.** Notably, Raman shifting rate was calculated with respect to the tensile/compressive strain along the in-plane direction of FeTe. Both $E_{2g}$ and $A_{1g}$ Raman modes of $MoS_2$,[46] $WS_2$,[47] $WSe_2$[48] and $VS_2$[49] presented redshift under uniaxial strain. The $B_{2g}$ and $A_g^1$ Raman modes of b-As showed blueshift and redshift, respectively, when the uniaxial strain applied along its armchair direction.[50] In contrast, both in-plane and out-of-plane Raman modes of $Cr_2S_3$,[33] GeSe,[51] $TaS_2$[52] and FeTe (this work) exhibited blueshift under strain.

Furthermore, the linear relationship was acquired according to the fitted phonon frequencies of 2D FeTe nanoflakes with the variable thickness under uniaxial strain. As shown in Figure 5, the Raman slopes of in-plane $E_g$ mode and out-of-plane $A_{1g}$ mode of FeTe exhibited strong thickness dependence under both compressive (green) and tensile (orange) strain. Raman slope $K$ is defined as $d\omega/d\varepsilon$, where $\omega$ and $\varepsilon$ are the Raman shift and the applied strain, respectively. In Figure 5a, the representative linear rates of $E_g$ mode of $t$-FeTe nanoflakes under tensile strain decreased from 5.6 to 0.6 cm$^{-1}$%$^{-1}$, corresponding to the growing thickness from 9.3 to 41.8 nm. For most typical 2D materials, the interlayer sliding could appear for thicker nanoflakes, leading to the remarkably reduced slopes.[61] Under the compressive strain, the $K$ of $E_g$ mode varied from -4.4 to -1.5 cm$^{-1}$%$^{-1}$ as the thickness of $t$-FeTe increased. In Figure 5c, the linear rates of $A_{1g}$ mode of $t$-FeTe nanoflakes under tensile and compressive strain exhibit the comparable thickness-dependent relationships as that of $E_g$ mode. The absolute values of $K$ of $A_{1g}$ mode under uniaxial strain decreased as its growing thicknesses. In comparison, the slope $K$ of $E_g$ mode of $h$-FeTe under tensile strain decreased from 7.0 to 3.9 cm$^{-1}$%$^{-1}$, corresponding to the varied thickness from 11.8 to 37.8 nm in Figure 5b. The slope $K$ under compressive strain varied from -3.9 to -2.0 cm$^{-1}$ %$^{-1}$ as the thickness of $h$-FeTe nanoflakes increasing. Interestingly, the slope $K$ of $A_{1g}$ mode of $h$-FeTe under tensile and compressive strain presents an opposite thickness- dependence, as shown in Figure 5d.

On the other hand, the thickness-dependent linear rates of two Raman modes of $h$-FeTe were obviously weaker than that of $t$-FeTe nanoflakes, which can be attributed to their lattice difference along $z$ axis. The interlayer coupling of the layered $t$-FeTe is relatively weaker than the covalent bonds of non-layered $h$-FeTe.[62] The tensile strain effect on 2D FeTe nanoflakes should be more significant than compressive strain, because of its higher loading efficiency in the utilized system. Furthermore, for FeTe nanoflakes with thickness over ~25 nm, their slope $K$ almost remained insensitive to thickness, and gradually approached to the bulk limit. In principle, the Raman shifting rate with strain can be related to the force constant of this specific phonon vibration, and partly reflect the elastic modulus of FeTe, according to dynamic theory of crystal lattices.[63] The force constant reasonably decreased as the thickness of FeTe before the emergent plateau, indicating the gradually reduced Young's modulus of FeTe along its in plane direction, well consistent with our previous results.[64]

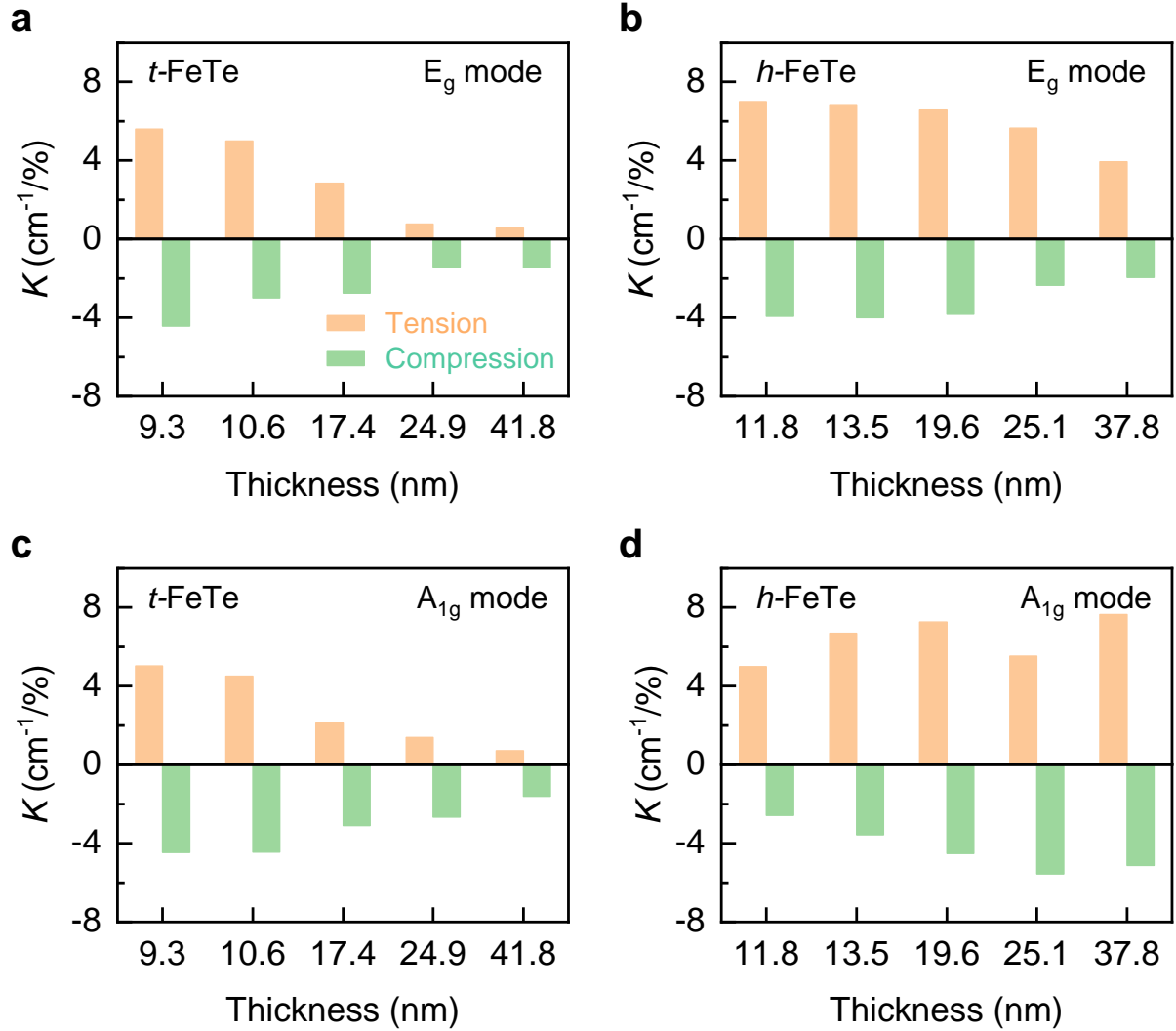

**Figure 5. The slopes of E$_g$ and A$_{1g}$ Raman shift under uniaxial strain with different thicknesses of 2D FeTe. a)** The obtained E$_g$ Raman slopes of *t*-FeTe as a function of its thickness. **b)** The E$_g$ Raman slopes of *h*-FeTe with variable thicknesses. **c)** The exacted A$_{1g}$ Raman slopes of *t*-FeTe as a function of its thickness. **d)** The A$_{1g}$ Raman slopes of *h*-FeTe with variable thicknesses.

**CONCLUSION**

In conclusion, we explored the abnormal lattice symmetry-dependent vibrational properties of 2D FeTe by the uniaxial strain engineering approach. The *in situ* characterizations of Raman spectroscopy confirmed that both *t*- and *h*-FeTe kept their structural stability under uniaxial strain up to 0.4%. The in-plane $E_g$ and out-of-plane $A_{1g}$ Raman modes were greatly sensitive to the external strain along its in-plane direction, corresponding to the hardened (softened) behaviors under tensile (compressive) strains owing to its strong spin-phonon coupling. Moreover, the obtained Raman shifting rates of both $E_g$ and $A_{1g}$ modes presented a clear thickness-dependence. We hope these results could give assistance to comprehensively understand the optical and mechanical properties of 2D magnetic FeTe and further develop its applications in numerous fields.

## EXPERIMENTAL METHODS

**Preparation and Characterization of 2D FeTe Nanoflakes**

An atmospheric pressure CVD approach was used to synthesize high-quality, large-area, and uniform 2D FeTe single crystals with two different phases on mica substrate. The detailed procedure for the sample preparation can be found in our previous work.[29] 2D FeTe nanosheets were transferred from mica surface onto the center of PET surface by a polystyrene-assisted method as established in our previous researches.[14,50] The typical dimension of PET substrate was 5 mm (width) ×50 mm (length), and about ~200 μm thick. Importantly, a thin PMMA film was coated to essentially prevent any oxidation and slippage of FeTe nanoflakes. AFM (Dimension FastScan, Bruker) characterizations were performed with tapping mode to accurately determine the thickness of 2D FeTe nanosheets before the bending test. For HRTEM measurements (Tecnai TF-20), 2D FeTe nanosheets were transferred from mica substrate onto TEM grids by using the non-polar polystyrene-assistant method, and the detailed procedure can be found elsewhere. The accelerating voltage was as low as 80 kV.

***In Situ* Raman Measurements**

*In situ* Raman measurements under strain was carried out by using our home-made Raman system, equipped with iHR550 spectrometer and 633 nm laser as the excitation source. The laser power was as low as tens of μW to exclude the potential overheating effect. All Raman characterizations under uniaxial strain were performed with a 50× objective lens and a grating of 1800 grooves mm$^{-1}$. A home-made four-point bending apparatus was used to apply the tunable and uniform strain along in-plane direction of 2D FeTe, as shown in Figure S1. In final, the versatile straining setup can effectively realize the tensile and compressive states of 2D nanoflakes by flipping the entire PET substrate over.

## ASSOCIATED CONTENT

**Supporting Information**

The Supporting Information is available free of charge and can be found online.

Schematics for the uniaxial strain setup and strain calculation method of our bending apparatus; Atomic force microscopy characterizations of pre-transferred 2D FeTe nanosheets with different thicknesses on PET substrate; Strain-dependent Raman spectrum of $t$-FeTe and $h$-FeTe with various thicknesses.

**Notes**

The authors declare no competing financial interests.

**Data Availability**

All data related to this study are available from the corresponding author on reasonable request.

**Author Contributions**

Y.C. and W.H. conceived this research project and designed the experiments. W.H. and L.Y. developed and established the four-point bending setup for applying the controlled strain. W.H., T.Z., and P.Z. carried out the Raman measurements. M.C. and J.S. prepared the FeTe samples via chemical vapor deposition approach. H.W. and Y.C. wrote the manuscript with the essential input of other co-authors. All authors have given approval of the final manuscript.

**Acknowledgements**

This work was financially supported by the National Natural Science Foundation of China (Grants 52072032 and 12090031), and the 173 JCJQ Program (Grant 2021-JCJQ-JJ-0159).